\documentclass[aps,prl,showpacs,twocolumn]{revtex4}

\usepackage{amssymb}
\usepackage{epsfig}
\usepackage{graphicx}
\usepackage{amsmath}
\usepackage{array,color}

\begin{document}

\title{ Interacting Dirac Fermions on Honeycomb Lattice}

\author{Wei Wu$^{1}$, Yao-Hua Chen$^1$, Hong-Shuai Tao$^1$, Ning-Hua Tong$^2$, and Wu-Ming Liu$^1$}
\affiliation{
$^1$Beijing National Laboratory for Condensed Matter Physics, Institute of Physics, Chinese Academy of Sciences, Beijing 100190, China \\
$^2$Department of Physics, Renmin University of China, Beijing 100872, China
}

\date{\today}

\begin{abstract}
 We investigate the interacting Dirac fermions on honeycomb lattice by cluster dynamical mean-field theory
 (CDMFT) combined with continuous time quantum Monte Carlo simulation (CTQMC).
 A novel scenario for the semimetal-Mott
insulator transition of the interacting Dirac fermions is
found beyond the previous DMFT studies.
We demonstrate that the non-local spatial correlations play a vital role in the Mott transition on the
honeycomb lattice. We also elaborate the experimental protocol to observe
this phase transition by the ultracold atoms on optical
honeycomb lattice.

\end{abstract}


\maketitle

The honeycomb lattice systems have been intensively studied
recently, where a number of exotic phenomena are found in both
experiment and theory, such as the single-layer graphene
\cite{Novoselov, Kane, YYzhang}, the topological Mott insulator
\cite{Raghu} and the recently found quantum spin liquid \cite{Meng}.
The electrons on honeycomb lattice can be described by a
\textquotedblleft relativistic\textquotedblright  massless
Dirac-fermion model \cite{Semenoff}, which is characterized by the
linear low-energy dispersion relation
$E(\textbf{k})={\pm}v_{F}|\textbf{k}|$. For half-filling case, the
Fermi surface of honeycomb lattice is reduced to isolated points
at the corners of Brillouin zone, hence the density of states
vanishes at the Fermi level, leading
  to the so called semimetal. When the repulsive interaction between
the fermions is turned on, one can expect this semimetal to be driven into
 a gapped insulator by the sufficient strong interaction, in other
words, a semimetal-Mott insulator transition happens. The Dirac
fermion on honeycomb lattice is massless, hence the prevalent Fermi liquid
theory is not proper here. Moreover, the semimetal has a small
spectral weight near the Fermi surface, thus the low-energy
many-body scattering effects would be quite different from those of
the usual metallic systems. As a consequence of these anomalous
natures, a new scenario is desirable for the description of Mott
transition of the interacting Dirac fermions. There have been
various theoretical studies of phase transitions on the honeycomb
lattice, however up to date no comprehensive conclusion has been
achieved \cite{Sorella,Honerkamp,Martelo}.

 Due to the efficient description
of the quantum fluctuations, the DMFT
\cite{Metzner,Georges,GKKR,Bulla} and its cluster extensions
\cite{Kotliar,Jarrel,Tong} have made substantial progress in the
field of Mott transition in the past decades. The recent DMFT
studies on honeycomb lattice find a first-order Mott transition and
the renormalization of the Fermi velocity $v_{F}$ induced by the
interaction \cite{Tran,Jafari}. Nevertheless, the DMFT ignores
the non-local correlations. It is efficient only in physical systems
with high dimensions or large coordination numbers \cite{Metzner}.
The CDMFT method \cite{Kotliar}, as a cluster extension of DMFT,
effectually incorporates the spatial correlations by mapping the
lattice problem into a self-consistently embedded cluster rather
than a single site in DMFT. For low-dimensional systems, the quantum
fluctuations are much stronger than in the higher dimensions, hence
the CDMFT method is more precise than the DMFT here. To our knowledge,
the DMFT calculations on square lattice basically agree with the
CDMFT. The local approximation of DMFT captures well the main
characteristics of the first-order Mott transition \cite{Park}.
However, in the 1D chain lattice, the ignorance of non-local correlations within DMFT leads to results with
severe errors \cite{Bolech}. The coordination number of honeycomb
lattice is three which is between those of 2D square lattice and 1D
chain lattice. Therefore, it is interesting to see whether the
non-local correlations can significantly affect the behaviors of the
electrons on honeycomb lattice and whether the CDMFT can provide
novel results beyond the DMFT.

In this Letter, we combine the CDMFT method with CTQMC simulation \cite{Rubtsov}, which is employed as impurity solver,
 to investigate the interacting Dirac fermions. A second order Mott transition is suggested. We also find
a number of novel features of the Dirac
 fermions, including the quasi-particles with anomalous long lifetime and the
invariable density of low-energy excitations which is highly relevant to the non-local correlations.
Our results are distinct
from those of the previous DMFT studies \cite{Jafari, Tran}, which indicate the
non-local correlations are indeed crucial in the honeycomb lattice.
\begin{figure}[!t]
\begin{center}
\scalebox{0.45}[0.45]{\includegraphics*[50pt,590pt][580pt,790pt]{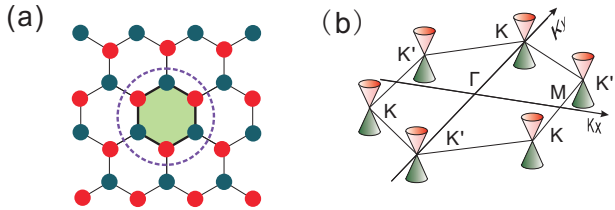}}
\end{center}
\caption{(Color online). a) Illustration of the honeycomb lattice.
The dashed line sketches the 6-site cluster scheme
 we explore in this work. b) The first Brillouin zone of the honeycomb lattice. The linear low-energy dispersion relation
  displays conical shapes near the Fermi level. }
\end{figure}

We consider the standard Hubbard model on the honeycomb lattice defined by
the Hamiltonian,
\begin{equation}
H = -t \sum_{<i,j>,\sigma} c^{\dagger}_{i\sigma} c^{\null}_{j\sigma}
+ U \sum_{i} n_{i\uparrow}n_{i\downarrow} + \mu \sum_{i} c^{\dagger}_{i\sigma} c^{\null}_{i\sigma},
\label{hubbard}
\end{equation}
where $c^{\dagger}_{i\sigma}$ and
$c_{i\sigma}$ are the creation and
the annihilation operator of fermions with site index $i$
and spin index $\sigma$, $n_{i\sigma}=c^{\dagger}_{i\sigma} c_{i\sigma}$ is the density operator. $U$ is the on-site repulsion,
and $\mu$ is the chemical potential. The nearest neighbor hopping amplitude $t$ $(t\!>\!0)$ determines the
bare dispersion of the honeycomb lattice
$E(\textbf{k})={\pm}t\sqrt{3+2cos(k_{y}a)+4cos(\sqrt{3}k_{x}a/2)cos(k_{y}a/2)}$, where $a$ is the distance between the two adjacent sites, the
plus and minus sign respectively denotes the upper ($\pi$) and lower ($\pi^{*}$) band.
 At low energies, the dispersion relation can be approximated as a linear one
$E(\textbf{k})={\pm}v_{F}|\textbf{k}|$, which can be described by the Dirac cones as shown in Fig. 1b. We take $t$ as the energy unit throught
out this paper.

The CDMFT method has been successfully applied to the
studies on Mott transition of the Hubbard model for various systems
\cite{Parcollet,Ohashi, Ohashi2006}. We use CDMFT to map the
original honeycomb lattice onto a 6-site effective cluster
embedded in a self-consistent medium. The effective cluster model is
obtained via an iterative procedure which can be started
with a initial guess of the cluster self-energy $\Sigma(i\omega)$. The effective
medium represented by the Weiss function $g(i\omega)$ is determined by the cluster self-energy $\Sigma(i\omega)$ via the coarse-grained Dyson equation,
\begin{equation}
{g}^{-1}(i\omega)=\big( \sum_\textbf{K}\frac{1}{{i\omega}+\mu-{t}(\textbf{K})-\Sigma({i\omega})} \big)^{-1} + \Sigma({i\omega}),
\end{equation}
where $t(\textbf{K})$ is the Fourier-transformed $6\times6$ hopping
matrix with wavevector $\textbf{K}$ in the cluster reduced Brillouin
zone of the superlattice, and $\mu$ is the chemical potential. Since
the effective cluster model has been defined by $g(i\omega)$,
 we can employ the numerical methods \cite{Rubtsov,Hirsch,Werner, SWZhang} as impurity solver to calculate the cluster Green's function $G(i\omega)$.
By using Dyson equation $\Sigma(i\omega)$=$g^{-1}(i\omega) -
G^{-1}(i\omega)$, we then recalculate the cluster self-energy
$\Sigma(i\omega)$ to close the self-consistent iterative loop. This
CDMFT loop is repeatedly iterated until the numerical convergence
has been achieved. In the present work, we use the numerically exact
CTQMC simulation as impurity solver and take about $7\times10^7$ QMC
sweeps for each CDMFT loop.

We first study the density of states (DOS) derived from the imaginary
time Green's function $G(\tau)$ by
 using maximum entropy method \cite{Gubernatis}.
The DOS for several values of on-site interaction $U$ at temperature
$T/t=0.05$ are presented in Fig. 2. When the interaction is
increased, the two quasi-particle peaks above and blow Fermi level
shift to the Fermi surface.
 The spectral weight is continuously transferred to the higher energy states, and eventually
 a gap opens at the Mott transition critical value $U_{c}/t\sim3.7$.
However, it is surprisedly noted that the DOS near Fermi level is
independent of the interaction strength $U$ until the Mott transition
happens. The quasi-particles in semimetal regime are always
analogous to the non-interacting Dirac gas, indicating no
renormalization of the Fermi velocity. This result is completely different from
the previous DMFT studies \cite{Tran,Jafari},
where the results suggest the DOS near Fermi level increases
as $U$ increases in the semimetal region,
hence the Fermi velocity $v_{F}$ is renormalized by the on-site
interaction.
 We argue that the non-local correlations
which are absent in DMFT, but included effectively by CDMFT significantly reduce the low-energy spectral weight.
\begin{figure}[!t]
\begin{center}
\scalebox{0.45}[0.45]{\includegraphics*[75pt,170pt][536pt,550pt]{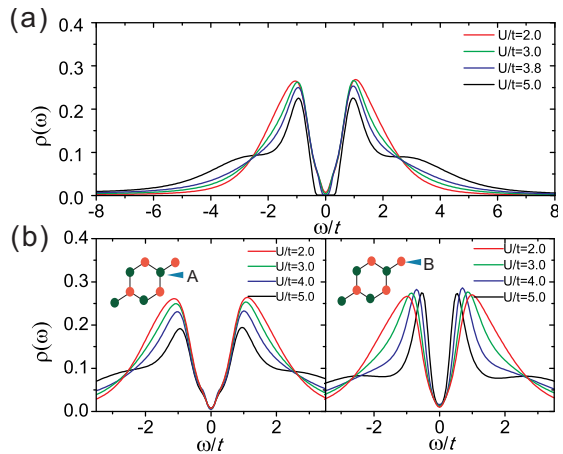}}
\end{center}
\caption{(Color online). The density of states for different
interaction strength $U$ at temperature $T/t=0.05$. a) Within 6-site
CDMFT, the DOS near Fermi level is invariable until the Mott
transition happens, which denotes a constant Fermi velocity $v_{F}$ in the semimetal regime. b) The 8-site cluster results. Insert: geometry
of the 8-site cluster. Left panel: the low-energy DOS of central site
A is invariable with the increasing of $U$,
which agrees with the 6-site CDMFT result. Right panel: the low-energy DOS of
boundary site B increases as $U$
increases, resembling the DMFT result. Note that
the Mott transition critical value $U_{c}$ of the 8-site cluster is
much larger than that of the 6-site cluster and close to the DMFT
result. This is due to the boundary effects. }

\end{figure}
\begin{figure}[!t]
\begin{center}
\scalebox{0.5}[0.5]{\includegraphics*[134pt,490pt][495pt,700pt]{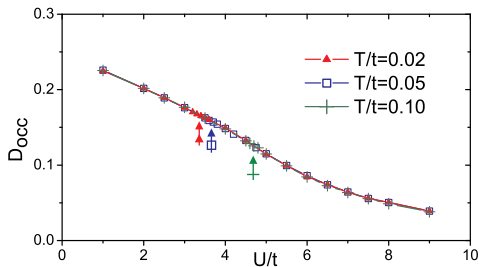}}
\end{center}
\caption{(Color online). Double occupancy $D_{occ}$ as a function of
interaction $U$ at different temperatures. The double occupancy is
insensitive to temperature for both weak- and strong-coupling
regime, hence the three curves for temperature $T/t=0.02,0.05,0.10$ strictly
superpose each other. The double occupancies decrease smoothly as
$U$ increases, indicating second-order Mott transitions. The
transition points for different temperatures are marked by the
arrows. }
\end{figure}

To clarify the spectral reduction effect of the non-local
correlations, we study an 8-site cluster which has two boundary
sites with only one neighbor site in the cluster. The geometry of the 8-site cluster is shown in inserts of Fig. 2b. It is apparent that the
reduction of the coordination number
 subdues the non-local correlations for boundary sites \cite{Biroli},
 therefore the characters of the boundary sites may be distinguished from those of
the central sites. As shown in Fig. 2b, the central site shows
invariable low-energy DOS coinciding with the 6-site
CDMFT study. However, the result of boundary site suggests
low-energy DOS increasing as $U$ increases, which is
just a characteristic observed by the DMFT. The central
and boundary sites are studied in one cluster, this disparity can only be attributed to the non-local correlations. We
therefore conclude that the DMFT incorrectly suggests an increasing low-energy DOS
 due to the ignorance of the non-local correlations.

We further investigate the double occupancy defined by the first
derivative of the free energy $F$,
$D_{occ}=\frac{\partial{F}}{\partial{U}}=\langle
n_{i\uparrow}n_{i\downarrow}\rangle$ as a function of $U$ for
different temperatures. As shown in Fig. 3, we find that when the
temperature is lower than $T_{c}/t\sim0.1$, the $D_{occ}$ is
independent of $T$ in both semimetallic and insulating phase. Unlike
the DMFT result \cite{Tran} or CDMFT studies on triangular lattice
\cite{Parcollet} and kagom\'e  lattice \cite{Ohashi2006}, where the
double occupancies $D_{occ}$ are insensitive to the temperature in
only insulating phase. It is also remarkable that
when $U$ increases, the $D_{occ}$ curves decreases
 smoothly on the entire domain of $U$  which indicates second-order phase transitions.
There is no hysteresis phenomenon observed in our calculations,
which also signals continuous phase transitions. This feature is
expected, considering the fact that regardless of the on-site
interaction strength, the upper and lower Hubbard band of honeycomb
lattice always touch each other in only Dirac points in the
semimetallic phase. Therefore the order parameter $D_{occ}$
decreases continuously even when the gap between the two Hubbard
bands opens.
\begin{figure}[!t]
\begin{center}
\scalebox{0.45}[0.40]{\includegraphics*[20pt,26pt][430pt,330pt]{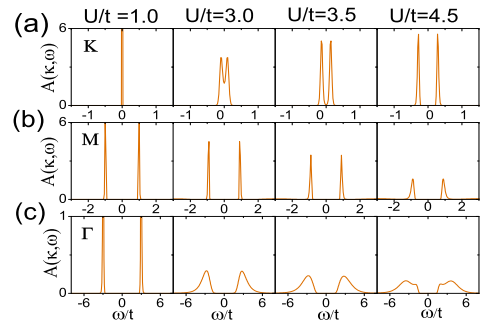}}
\end{center}
\caption{(Color online) The spectral functions $A(k,\omega)$ for a)
$K$ point. We can see the pseudogap at Fermi level developing as the
interaction strength $U$ increases. b) $M$ point. The peaks of
$A(k,\omega)$ are very sharp even in the strong coupling regime,
which indicates well-defined quasi-particles. c) $\Gamma$ point. Far away
from the Fermi level, the quasi-particle peaks are rapidly broadened
by the interaction. }
\end{figure}

 Fig. 4 shows the spectral functions $A(k,\omega)$ at $K$ point $k=(0,4{\pi}/3a)$ , $M$ point $k=(2\sqrt3\pi/3a, 0)$ and $\Gamma$ point $k=(0,0)$.
  For the $K$ point (Dirac point), as the interaction $U$ is increased, $U/t=3.0$ for example, the quasi-particle peak at Fermi surface ($\omega=0$) is strongly
 suppressed by the antiferromagnetic correlations, and a pseudogap develops. When $U$ reaches the critical value
  $U_{c}/t\sim3.7$ for the phase transition,
the pseudogap eventually engulfs the whole Fermi surface and the
system turns into Mott insulator. The observation of pseudogap on
the Dirac point in the transition regime indicates the
antiferromagnetic correlations are the origin of the Mott phase transition.
Note that while the quasi-particle peaks at $\Gamma$ point are
rapidly broadened by the on-site interaction, the spectral functions
$A(k,\omega)$ at the $K$ and $M$ point have sharp peaks even in the
strongly coupling regime, indicating a long quasi-particle lifetime.
This is related to the fact that there are few electrons near the
Dirac points, thus the many-particle scattering effects are not
prominent for the $K$ and $M$ points.

Although the local density of states
$\rho(\omega)=\sum_{k}A(k,\omega)$ is independent of the
interaction for small energy $\omega$, the states are
redistributing in the momentum space when the interaction increases.
As shown in Fig. 5a, the distribution of quasi-particles with energy
$\omega/t=0.4$ near a Dirac point $(2\sqrt{3}\pi/3,2\pi/3)$ becames
isotropic in different directions when $U$ increases and finally the
spectra averagely distribute in the whole ring. The Fermi surface
denoted by $A(k,\omega/t=0)$ is also presented.
The first Brillouin zone of honeycomb lattice has three $K$ and three $K^{'}$ points located at the corners
of the hexagon, as shown in Fig. 1b. Due to the particle-hole
symmetry, the six Dirac points are actually all equivalent in our study. As a result, the Mott phase transition
happens isochronously on the six Dirac points. Fig. 5b also shows that
the locations of Dirac points are steady in momentum space with the increasing of interaction $U$.

\begin{figure}[!t]
\begin{center}
\scalebox{0.45}[0.45]{\includegraphics*[60pt,270pt][450pt,535pt]{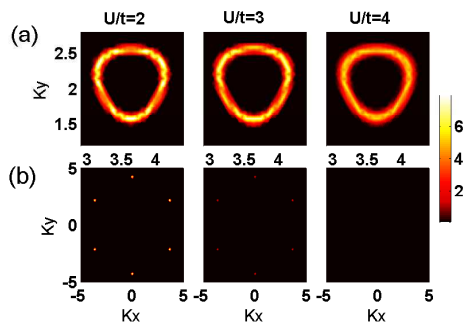}}
\end{center}
\caption{(Color online) The distribution of $A(k,\omega)$ in
momentum space for different interactions. a) The distribution of
$A(k,\omega/t=0.4)$ near a Dirac point $(2\sqrt{3}\pi/3,2\pi/3)$
becomes isotropic in different directions when $U$ increases. b)
$A(k,\omega/t=0)$ depicts the Fermi surface of honeycomb lattice
reduced to six Fermi points.}
\end{figure}
Finally, we present the phase diagram of the interacting Dirac fermions on
honeycomb lattice. The diagram is separated into semimetallic and
insulating phases by the second-order Mott transition line. The
critical value $U_{c}$ decreases as temperature $T$ decreases, and
 a finite $U_{c}/t\sim3.3 $ for zero temperature phase transition is suggested, which is quite agree with
 the large scale quantum Monte carlo result $U_{c}/t\sim3.5 $ \cite{Meng}. Comparing to the DMFT
 result $U_{c}/t\sim10$ \cite{Tran}, the Mott transition critical value $U_{c}$ within CDMFT is much
smaller, and more reasonable according to the previous studies
\cite{Sorella,Honerkamp,Martelo}.

On the experiment side, due to the high controllability and
clearness, the interacting ultracold atoms on the optical honeycomb
lattice may be an ideal tool for the verification of our results \cite{Greiner, Gemelke, Dudarev}.
The optical honeycomb lattice can be produced by three intersected
standing-wave lasers. Each of the standing wave is formed by two
detuned laser beams intersecting at an angle of $70.4^{\circ}$. The
ultracold fermion atoms like the $^{40}K$ is then loaded in the
optical lattice \cite{Duan}. The on-site interaction between the
fermion atoms can be adjusted from zero to strong coupling limits by
the Feshbach resonance. Using absorption imagining technology, the
crucial physical quantities such as the double occupancy and the
Fermi surface can be obtained to analyze the system \cite{Jordans}.

In summary, we study the Mott transition of the interacting Dirac
fermions on honeycomb lattice by CDMFT method. It is found
that the low-energy density of states is independent of the
interaction strength in the semimetal regime, hence the Fermi
velocity always equals to the one of non-interacting Dirac sea. As
the interaction increases, the system enters Mott insulator
via a second-order phase transition accompanied by a pseudogap on
the Fermi surface. This suggests the Mott transition is dominated by
the Slater mechanism. It is also shown that the non-local
correlations significantly reduce the low-energy density of states of the
honeycomb lattice. We argue that because of lacking non-local
correlations, the DMFT method is inadequate in the study of strongly
correlated fermions on honeycomb lattice. Our study may
 contribute to the comprehension of the interaction
driven Mott transition and the recently found quantum spin liquid on
honeycomb lattice \cite{Meng}.

\begin{figure}[!htb]
\begin{center}
\scalebox{0.35}[0.35]{\includegraphics*[30pt,30pt][510pt,330pt]{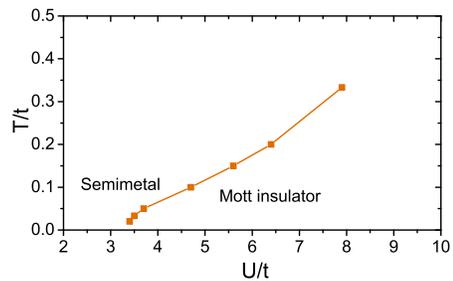}}
\end{center}
\caption{(Color online) Phase diagram of the interacting Dirac fermions on honeycomb
lattice. The line separating the semimetal and Mott insulator marks
a sencond-order phase transition. }
\end{figure}

This work was supported by NSFC under grants Nos. 10874235,
10934010, 60978019, the NKBRSFC under grants Nos. 2006CB921400, 2007CB925004, 2009CB930701 and 2010CB922904.
The numerical computations was done at the supercomputing center of CAS.

\end{document}